\let\accentvec\vec 
\RequirePackage{fix-cm}
\documentclass[smallcondensed]{svjour3}     
\let\vec\accentvec
\usepackage{graphicx}
\usepackage{dcolumn}
\usepackage{bm}
\usepackage{amssymb,amsfonts,amsmath}
\usepackage{makecell}
\usepackage{braket} 
\usepackage{enumerate}
\usepackage{soul}
\usepackage{xcolor}
\usepackage{color}
\usepackage{tcolorbox}

\soulregister\cite7 
\soulregister\citep7 
\soulregister\citet7 
\soulregister\ref7 
\soulregister\pageref7 
\spnewtheorem{myDefinition}{Definition}{\bf}{\rm}
\spnewtheorem{myLemma}{Lemma}{\bf}{\rm}
\spnewtheorem{myTheorem}{Theorem}{\bf}{\rm}

\journalname{Quantum Information Processing}
\begin{document}
\title{One Dimensional Quantum Walks with Two-step Memory
}


\author{Qing Zhou         \and
        Songfeng Lu 
}
\institute{Qing Zhou \at
              School of Computer Science and Technology, Huazhong University of Science and Technology, Wuhan 430074, China \\
           \and
           Songfeng Lu \at
              School of Computer Science and Technology, Huazhong University of Science and Technology, Wuhan 430074, China \\
            Corresponding author,   \email{lusongfeng@hotmail.com}             \\
}
\date{Received: date / Accepted: date}

\maketitle

\begin{abstract}
In this paper we investigate one dimensional quantum walks with two-step memory, which can be viewed as an extension of quantum walks with one-step memory. We develop a general formula for the amplitudes of the two-step-memory walk with Hadamard coin by using path integral approach, and numerically simulate its process. The simulation shows that the probability distribution of this new walk is different from that of the Hadamard quantum walk with one-step memory, while it presents some similarities with that of the normal Hadamard quantum walk without memory.
\end{abstract}

\section{Introduction}\label{sec:1}
Quantum walks are quantum analogues of classical random walks, which usually exhibit different features compared with their classical counterparts due to quantum interference phenomena~\cite{Ambainis2001}, and they have many interesting applications in computer science~\cite{Xue2019,Feng2019,Marsh2019,Vlachou2018,Yang2019}. As a specific category of Markov processes, classical and quantum random walks are normally memoryless---at each time step, the next move of the walker only depends on the current state and is irrelevant to the historical states. This condition (known as Markovian property) does not hold in quantum walks with memory---the order-2 quantum walks, in which the walker moves conditional on both its current and previous positions~\cite{Gettric2010QW1M}---in the sense that the standard system (including the coin and the current position) considered in quantum walks are coupled to an additional subsystem indicating the last position of the particle (the ``environment''). On the other hand, if the system we consider is broadened to include the last position, then the purely unitary evolution of this system leads to Markovian dynamics, and such a model can be viewed as a multi-state quantum walk without memory~\cite{Konno2010Limit}. Such two perspectives on quantum walks with memory are inspired by Ref.~\cite{Kummar2015NonMarkov,Rivas2014NonMarkov,Hinarejos2014Chirality}.'' For clarity, we refer to order-2 quantum walks as quantum walks with one-step memory (QW1M), and refer to quantum walks where the next move is related to $n$ steps before the current state as quantum walks with $n$-step memory or order-$(n+1)$ walks. Additionally, ``standard'' quantum walks are referred to as quantum walks without memory (QW0M) or order-1 walks.

Here, we extend the historical steps that govern the next move in QW1M one step further, namely, to investigate one dimensional quantum walks with two-step memory (QW2M) both numerically and analytically. The numerical simulation shows that QW2M with Hadamard coin does not have the localization property presented in Hadamard QW1M (i.e., a high probability in the origin), while it exhibits a double-peaked shape in the probability distribution, which is similar to that of QW0M.

This paper is structured as follows. First, we give a detailed description of one-dimensional QW2M in Section~\ref{sec:2}. Then, we derive a general amplitude formula for the Hadamard QW2M in Section~\ref{sec:3}, which acts as a guideline of deducing each amplitude for this walk. The lengthy but precise amplitude expressions (along with derivations) are listed in the full version of this paper (in Appendix C)~\cite{Zhou2019QW2M}, whose correctness has been verified by numerical simulations. We illustrate and compare the probability distributions of QW2M, QW1M, and QW0M in Section~\ref{sec:4}, and finally conclude in Section~\ref{sec:5}. The proof of each lemma and theorem is presented in appendix~\ref{app:appA}.
\section{Quantum walks with two-step memory}
\label{sec:2}
The state space of one dimensional QW2M is spanned by the vectors of the form
\begin{equation}
\Ket{n_3,n_2,n_1,p}\label{eq:1},
\end{equation}
where $p$ (with $p\in\{0,1\}$) is the coin state, $n_1$ is the current position and $n_i$ $(i\in\{2,3\},\left|n_i-n_{i-1}\right|=1)$ is the position $i-1$ steps ago. According to the sign of $n_i-n_{i-1}$, the state $\Ket{n_3,n_2,n_1,p}$ (hereafter referred to as an \emph{original state}) can be converted into another type of state (hereafter referred to as a \emph{direction state}) of the form
\begin{equation}
\Ket{dr_2,dr_1,p}\label{eq:2},
\end{equation}
in which $dr_1$ is the direction of the last step, and $dr_2$ is the direction immediately before the last step. In other words, if $n_i-n_{i-1}=1$, then $dr_i=L$, and if $n_i-n_{i-1}=-1$, then $dr_i=R$.
Moreover, a direction state can be associated to a \emph{trend state} of the form
\begin{equation}
\Ket{trd,p}\label{eq:3},
\end{equation}
where $trd$ indicates consistency or inconsistency between $dr_1$ and $dr_2$: if $dr_1=dr_2$, then $trd=C$ ($C$ means consistent directions), and if $dr_1\ne dr_2$, then $trd=O$ ($O$ means opposite directions).

By considering 2-state quantum walks with memory as $m$-state ($m>2$) quantum walks without memory~\cite{Konno2010Limit}, the state vector $\Ket{n_3,n_2,n_1,p}$ can also be rewritten as
\begin{equation}
\Ket{n_1,(n_1\!-\!n_2\!+\!1)\!\cdot\!2^1\!+\!(n_2\!-\!n_3\!+\!1)\!\cdot\!2^0\!+\!p\!\cdot\!2^0}
=\Ket{n_1,2n_1\!-\!n_2\!-\!n_3\!+\!p\!+\!3}\label{eq:4}.
\end{equation}
Conversely, one can deduce $\Ket{n_3,n_2,n_1,p}$ from $\Ket{n_1,j}$ ($j\in\{0,1,\dots,7\}$): first, $p=j\text{ mod }2$; then, the identities $n_1-n_2=(j-j\text{ mod } 4)/2-1$ and $n_2-n_3=j\text{ mod }4-j\text{ mod }2-1$ give $n_2=n_1-(j-j\text{ mod }4)/2+1$ and $n_3=n_2-j\text{ mod }4+j\text{ mod }2+1=n_1-(j-j\text{ mod }4)/2-j\text{ mod }4+j\text{ mod }2+2$. Thus, there is a one-to-one correspondence between the original states for a fixed position $k$ and the eight computational basis states ranging from $\Ket{0}$ to $\Ket{7}$. For clarity, we present the correspondence among the original states, the direction states, the trend states, and the computational basis states (basis states for short) in Table~\ref{tab:table1}.
\begin{table}
\caption{\label{tab:table1}%
Correspondence among the four types of states
}
\begin{tabular}{llll}
\hline\noalign{\smallskip}
\textrm{Original states} & \textrm{Direction states} & \textrm{Trend states} & \textrm{Basis states}\\
\hline\noalign{\smallskip}
$\Ket{k+2,k+1,k,0}$ & $\Ket{L,L,0}$ & $\Ket{C,0}$ & $\Ket{0}$ \\
$\Ket{k+2,k+1,k,1}$ & $\Ket{L,L,1}$ & $\Ket{C,1}$ & $\Ket{1}$ \\
$\Ket{k,k+1,k,0}$ & $\Ket{R,L,0}$ & $\Ket{O,0}$ & $\Ket{2}$ \\
$\Ket{k,k+1,k,1}$ & $\Ket{R,L,1}$ & $\Ket{O,1}$ & $\Ket{3}$ \\
$\Ket{k,k-1,k,0}$ & $\Ket{L,R,0}$ & $\Ket{O,0}$ & $\Ket{4}$ \\
$\Ket{k,k-1,k,1}$ & $\Ket{L,R,1}$ & $\Ket{O,1}$ & $\Ket{5}$ \\
$\Ket{k-2,k-1,k,0}$ & $\Ket{R,R,0}$ & $\Ket{C,0}$ & $\Ket{6}$ \\
$\Ket{k-2,k-1,k,1}$ & $\Ket{R,R,1}$ & $\Ket{C,1}$ & $\Ket{7}$ \\
\hline\noalign{\smallskip}
\end{tabular}
\end{table}

Roughly speaking, the one-step time evolution for QW2M can be decomposed into three steps: first, apply a unitary transform on the coin state to get a new coin $p$; second, determine the direction of the next move based on $p$ and the former two directions; and finally, shift according to the new direction. The first and the third steps are respectively an ordinary coin flip and a normal shift operator, while the second step is an additional procedure compared with QW0M, which is a crucial part of the time evolution.
\subsection{Direction-determine transform}
For convenience, we define the direction-determine process based on the trend states: if the coin state is 0, reflect the trend; and if the coin state is 1, leave the trend alone. This process can be formulated as
\begin{equation}
\Ket{C,0}\to\Ket{O,0},\Ket{O,0}\to\Ket{C,0},
\Ket{C,1}\to\Ket{C,1},\Ket{O,1}\to\Ket{O,1},\label{eq:5}
\end{equation}
which can also be defined on the direction states equivalently:
\begin{eqnarray}
\Ket{L,L,0}\to\Ket{L,R,0},\Ket{L,L,1}\to\Ket{L,L,1},\nonumber\\
\Ket{R,L,0}\to\Ket{L,L,0},\Ket{R,L,1}\to\Ket{L,R,1},\nonumber\\
\Ket{L,R,0}\to\Ket{R,R,0},\Ket{L,R,1}\to\Ket{R,L,1},\nonumber\\
\Ket{R,R,0}\to\Ket{R,L,0},\Ket{R,R,1}\to\Ket{R,R,1}.\label{eq:6}
\end{eqnarray}
Here, each state on the left of an arrow is a result state after a coin flip transform. If the coin flip (say, a Hadamard transform) is included, then the combined action on $\Ket{L,L,0}$ gives $1/\sqrt{2}\left(\Ket{L,R,0}+\Ket{L,L,1}\right)$.

Specifically, the transform above manipulates a direction state $\Ket{dr_2,dr_1,p}$ in two steps. First, determine the next direction $d$ in the following way: if $p=0$ and $dr_1\ne dr_2$, then $d=dr_1$; if $p=0$ and $dr_1=dr_2$, then $d\ne dr_1$ (i.e., $d$ is opposite to $dr_1$); and if $p=1$, then $d=dr_2$. Second, update the direction state to $\Ket{dr_1,d,p}$. By substituting the direction states with the corresponding computational basis states, one can construct a unitary matrix for~(\ref{eq:6}), suggesting that the time evolution for QW2M is unitary.
\subsection{Initial conditions}
A quantum walk with two-step memory has to be initiated with a standard walk without memory (where the walker goes left if the flipped coin state is 0, and goes right if the flipped result is 1). For ease of analysis, we assume that the walker starts at the origin, and that it first moves to the position 1 (with a coin state 1), then go back to the origin (with a coin state 0). This initial condition creates the initial state $\Ket{0,1,0,0}$, and from there on we run the order-3 walk.
\subsection{The Hadamard walk}
In general, any 2-by-2 unitary coin operator together with an arbitrary unitary direction-determine transform lead to a valid quantum walk with two-step memory. For concreteness, we focus on the QW2M with Hadamard coin, whose direction for each step is determined by~(\ref{eq:6}). The first few steps of this walk are
\begin{eqnarray}
&&\Ket{0,1,0,0}\to 1/\sqrt{2}\left(\Ket{1,0,-1,0}+\Ket{1,0,1,1}\right)\\
\to&& 1/2(\Ket{0,-1,0,0}+\Ket{0,-1,-2,1}+\Ket{0,1,2,0}-\Ket{0,1,0,1})\\
\to&&1/(2\sqrt{2})(\Ket{-1,0,1,0}+\Ket{-1,0,-1,1}+\Ket{-1,-2,-1,0}-\Ket{-1,-2,-3,1}\nonumber\\
&&+\Ket{1,2,1,0}+\Ket{1,2,3,1}-\Ket{1,0,-1,0}+\Ket{1,0,1,1})\\
\to&&1/4(\Ket{0,1,0,0}+\Ket{0,1,2,1}+\Ket{0,-1,-2,0}-\Ket{0,-1,0,1}+\Ket{-2,-1,0,0}\nonumber\\
&&+\Ket{-2,-1,-2,1}-\Ket{-2,-3,-2,0}+\Ket{-2,-3,-4,1}+\Ket{2,1,0,0}\nonumber\\
&&+\Ket{2,1,2,1}+\Ket{2,3,2,0}-\Ket{2,3,4,1}-\Ket{0,-1,0,0}-\Ket{0,-1,-2,1}\nonumber\\
&&+\Ket{0,1,2,0}-\Ket{0,1,0,1})\\
\to&&1/(4\sqrt{2})(\underline{\Ket{1,0,-1,0}}+\Ket{1,0,1,1}+\Ket{1,2,1,0}-\Ket{1,2,3,1}+\Ket{-1,-2,-1,0}\nonumber\\
&&+\Ket{-1,-2,-3,1}-\Ket{-1,0,1,0}+\Ket{-1,0,-1,1}+\Ket{-1,0,-1,0}+\Ket{-1,0,1,1}\nonumber\\
&&+\Ket{-1,-2,-3,0}-\Ket{-1,-2,-1,1}-\Ket{-3,-2,-1,0}-\Ket{-3,-2,-3,1}\nonumber\\
&&+\Ket{-3,-4,-3,0}-\Ket{-3,-4,-5,1}+\Ket{1,0,1,0}+\Ket{1,0,-1,1}+\Ket{1,2,3,0}\nonumber\\
&&-\Ket{1,2,1,1}+\Ket{3,2,1,0}+\Ket{3,2,3,1}-\Ket{3,4,3,0}+\Ket{3,4,5,1}-\Ket{-1,0,1,0}\nonumber\\
&&-\Ket{-1,0,-1,1}-\Ket{-1,-2,-1,0}+\Ket{-1,-2,-3,1}+\Ket{1,2,1,0}+\Ket{1,2,3,1}\nonumber\\
&&\underline{-\Ket{1,0,-1,0}}+\Ket{1,0,1,1})\label{eq:11}
\end{eqnarray}
It can be seen that the interference first appears in the fifth step (e.g., in~(\ref{eq:11}), one can cancel the 1st and 31th terms, and add the 2nd and 32th terms, etc.), which differs both from QW0M and QW1M. In Hadamard QW0M, constructive and destructive interference terms appear in the third step, while in Hadamard QW1M, the interference appears in the fourth step.
%
\section{Path integral analysis of the Hadamard QW2M\label{sec:3}}
There are two general methods for analyzing the process of quantum walks: the path integral approach and the Schrodinger approach. Due to the difficulty of obtaining the eigenvalues of the time evolution operator (an 8-by-8 matrix) on the Fourier-transformed amplitude vector in QW2M, a Schrodinger analysis of this walk turns out to be intractable, hence we adopt the other method to derive the amplitudes. Although the expressions given by the path integral approach are long and opaque, they are exact for all times.

Let $a_j$ $(j\in\{0,1,\dots,7\})$ be the amplitude of $\Ket{k,j}$ (corresponds to the basis state $\Ket{j}$ in Table~\ref{tab:table1}), then the probability (when we measure) of finding the walker at position $k$ is $\begin{matrix} \sum_{j=0}^7 |a_j|^2 \end{matrix}$, where $a_j$ can be computed by summing over the signed amplitudes of the different paths leading to $\Ket{k,j}$, and then divide by a global coefficient dependent on the number of steps.

Since each path in a one-dimensional walk can be viewed as a sequence of left ($L$) and right ($R$) moves, we refer to a path in the Hadamard QW2M as a \emph{direction sequence}, which is an ordered list made of $L$'s and $R$'s. Recall that the initial state of our order-3 walk is $\Ket{0,1,0,0}$, so all direction sequences in this walk begin with $RL$. Let $N_L$ (respectively $N_R$) be the number of $L$'s (respectively $R$'s) in a direction sequence leading to position $k$ after $n+2$ steps (the first two steps are of the order-1 walk, and the remaining $n$ steps are of the order-3 walk), then we have
\begin{equation}
N_L+N_R=n+2,N_R-N_L=k.
\end{equation}

In outline, the amplitude of a specific final state can be calculated in four steps: 1) seek out the segment patterns that will give phase contributions in a general direction sequence, then derive a phase expression $(-1)^x$ for this sequence; 2) determine the exact range of the exponent $x$ of the phase; 3) count the number of direction sequences for a given value of $x$; and 4) sum over the signed counts of direction sequences that correspond to all possible values of $x$ and that lead to the final state. Before investigating, we first introduce some notions that will help to derive the amplitudes.
\begin{myDefinition}[Clusters, multi-element clusters, singular clusters, isolated marginal clusters, non-singular clusters]\label{def:def1}
In a direction sequence, a \emph{cluster} is a direction segment that consists of consecutive $L$'s or consecutive $R$'s, called an $L$ \emph{cluster} or an $R$ \emph{cluster}; $L$ clusters and $R$ clusters appear alternately. A \emph{multi-element cluster} contains multiple elements, while a \emph{singular} or an \emph{isolated marginal cluster} contains a single element. The difference between the two types of clusters of size one is, a singular cluster is bordered by another cluster on either side, while an isolated marginal cluster is at the left or the right margin of the direction sequence. \emph{Non-singular clusters} comprise multi-element clusters and isolated marginal clusters.
\end{myDefinition}
\begin{myDefinition}[Cluster mask, $L$ cluster mask, $S$ and $M$ groups]\label{def:def2}
A \emph{cluster mask} is a simplified form of a direction sequence, which is obtained by replacing each singular $L$ (or $R$) cluster, each multi-element $L$ (or $R$) cluster, and each isolated marginal $L$ (or $R$) cluster with an $S$ (or $\overline{S}$), an $M$ (or $\overline{M}$) and an $I$ (or $\overline{I}$), respectively. Eliminating $\overline{S}$'s, $\overline{M}$'s and $\overline{I}$'s from a cluster mask gives an $L$ \emph{cluster mask}. An $S$ (respectively $M$) \emph{group} is a string of consecutive $S$'s (respectively $M$'s) in an $L$ cluster mask; $M$ groups and $S$ groups appear alternately.
\end{myDefinition}

As an example of the definitions given above, the direction sequence $drs=LRRLLLRRLRLRRRLLLLR$ contains one isolated marginal $L$ cluster, one isolated marginal $R$ cluster, two singular $L$ clusters, one singular $R$ cluster, two multi-element $L$ clusters, and three multi-element $R$ clusters. The cluster mask for $drs$ is $I\overline{M}M\overline{M}S$-$\overline{S}S\overline{M}M\overline{I}$ , which is composed of an $L$ cluster mask $IMSSM$ and an $R$ cluster mask $\overline{M}$\,$\overline{M}$\,$\overline{S}$\,$\overline{M}$\,$\overline{I}$. In the $L$ cluster mask, there are two $M$ groups of size one and one $S$ group of size two. With these notions, the quantum phase of a general direction sequence can be investigated conveniently.
\subsection{The phase contributions of direction segments\label{sec:3.1}}
In a Hadamard walk, a phase factor of -1 occurs once two consecutive 1's appear among the results of coin flips, which corresponds to a pair of identical trends, $CC$ or $OO$, in QW2M. Also, it corresponds to one of the four direction segments---$LLLL$, $RRRR$, $LRLR$, and $RLRL$. In $LLLL$, the first two $L$'s indicate a consistent trend $C$, which continues due to a coin state 1 (thus a third $L$ appears), and such a trend continues again due to a second coin state 1 (thus a fourth $L$ appears). In $LRLR$, the first two directions $LR$ indicate an opposite trend $O$, which continues due to a coin state 1 (thus a second $L$ appears), and it continues again due to a second coin state 1 (thus a second $R$ appears). The other two segments, $RRRR$ and $RLRL$, induce a factor of -1 in a similar way. The phase contributions given by more general segments of the forms $LL\dots L$, $RR\dots R$, and $\dots LRLRL\dots$ (hereafter referred to as \emph{long $L$ clusters}, \emph{long $R$ clusters}, and \emph{alternate direction segments}, respectively) within a direction sequence are formulated as follows, which are proved in Appendix \ref{app:subsec1} and \ref{app:subsec2}.
\begin{myLemma}[The phase contribution of long clusters]\label{lem:lemma1}
Consider a direction sequence of size $n+2$, in which the numbers of $L$ clusters, $R$ clusters, $L$ clusters of size two, and $R$ clusters of size two are $C_L$, $C_R$, $C_L^2$, and $C_R^2$ respectively. Then, the total phase contribution given by all long clusters in this sequence is $(-1)^{n+C_L+C_R+C_L^2+C_R^2}$.
\end{myLemma}
\begin{myLemma}[The phase contribution of alternate direction segments]\label{lem:lemma2}
Consider a direction sequence in which there are $r$ singular $R$ clusters, each is bordered by a singular $L$ cluster and a non-singular $L$ cluster. Then, the total phase contribution given by all alternate direction segments in this sequence is $(-1)^r$.
\end{myLemma}
It can be concluded from the two lemmas that the phase of a direction sequence of size $n+2$ in the Hadamard QW2M is
\begin{equation}
(-1)^{n+C_L+C_R+C_L^2+C_R^2+r}.\label{eq:13}
\end{equation}
\subsection{The range of the exponent of the phase\label{sec:3.2}}
Since $L$ and $R$ clusters appear alternately in a direction sequence, the number $C_L$ of $L$ clusters and the number $C_R$ of $R$ clusters satisfy the relation $|C_L-C_R|\leq 1$. The exact difference between $C_L$ and $C_R$ only depends on the first and the last elements of the direction sequence: if the sequence is of the form $L\dots R$ or the form $R\dots L$, then $C_R=C_L$; if the sequence is of the form $L\dots L$, then $C_R=C_L-1$; and if the sequence is of the form $R\dots R$, then $C_R=C_L+1$. Since the order-3 walk begins with $RL$, and its last three directions are implied by the final quantum state, $C_R$ can be expressed in terms of $C_L$. Thus, the phase of a direction sequence with a known size in the order-3 walk is determined by $C_L^2$, $C_R^2$, and $r$, whose ranges are given by Lemmas~\ref{lem:lemma3} and \ref{lem:lemma4}.

In what follows, we introduce a binary parameter $t_2t_1t_0$ ($t_0,t_1,t_2\in\{0,1\}$) to indicate the type of the end of an $L$ cluster mask: if the mask ends with an $S$, then $t_2t_1t_0=010$; if the mask ends with $M$, then $t_2t_1t_0=001$; if the mask ends with $SI$, then $t_2t_1t_0=110$; if the mask end with $MI$, then $t_2t_1t_0=101$; and if the last item of the mask cannot be decided (e.g., the $L$ mask for the direction sequence of the form $RL\dots LRR$ or $RL\dots RRR$), then $t_2t_1t_0=011$. In addition, we adopt the Kronecker delta symbol $\delta_{x,y}$ and the indicator function
\begin{equation}
\mathbf{1}_{\mathbb{Z}^+}(x)=
\begin{cases}
1, & \text{if }x\in \mathbb{Z}^+\\
0, & \text{if }x\not\in \mathbb{Z}^+
\end{cases}
\end{equation}
to describe the related bounds.
\begin{myLemma}[The ranges of $C_L^2$ and $C_R^2$]\label{lem:lemma3}
Consider a direction sequence containing $N_L$ $L$'s, $N_R$ $R$'s, $C_L$ $L$ clusters, $C_R$ $R$ clusters, $C_L^1$ $L$ clusters of size one, and $C_R^1$ $R$ clusters of size one. Then, in this sequence, the number $C_L^2$ of $L$ clusters of size two satisfies either
\begin{equation*}
C_L^2=C_L-C_L^1=N_L-C_L\text{ or }\text{max}(0,3C_L-2C_L^1-N_L)\leq C_L^2\leq C_L-C_L^1-1,
\end{equation*}
and the number $C_R^2$ of $R$ clusters of size two satisfies either
\begin{equation*}
C_R^2=C_R-C_R^1=N_R-C_R\text{ or }\text{max}(0,3C_R-2C_R^1-N_R)\leq C_R^2\leq C_R-C_R^1-1.
\end{equation*}
\end{myLemma}
\begin{myLemma}[The range of $r$]\label{lem:lemma4}
Let $m_C=\{m_L,m_R\}$ be a cluster mask composed of an $R$ cluster mask $m_R$ and an $L$ cluster mask $m_L$ indicated by $t_2t_1t_0$, where $m_C$ begins with an $\overline{I}$, $m_L$ contains $C_L^1-t_2$ $S$'s, $t_2$ $I$ and $C_L-C_L^1$ $M$'s, $m_R$ contains $C_R-C_R^1$ $\overline{M}$'s, and the number of $\overline{S}$'s and $\overline{I}$'s in $m_R$ is $C_R^1$. Then, the number $g$ of $S$ and $M$ groups in $m_L$ satisfies
\begin{eqnarray}
&&g \ge 2-\delta_{C_L^1,t_2}-\delta_{C_L,C_L^1}\nonumber\\
&&\text{and}\\\label{eq:16}
&&g \leq 2\text{min}(C_L^1\!-\!t_2,C_L\!-\!C_L^1)+t_1\mathbf{1}_{\mathbb{Z}^+}(2C_L^1\!-\!C_L\!-\!t_2)
+t_0\mathbf{1}_{\mathbb{Z}^+}(C_L\!-\!2C_L^1\!+\!t_2),\nonumber
\end{eqnarray}
and the number $r$ of $\overline{S}$'s (each) bordered by an $S$ and an $M$ or by an $S$ and $I$ satisfies
\begin{eqnarray}
\text{max}(0,C_R^1-C_R+g+t_1\cdot t_2-1)\leq r \leq \text{min}(C_R^1-1,g+t_1\cdot t_2-1).
\end{eqnarray}
\end{myLemma}

The detailed derivations of the two lemmas in above are presented in Appendices~\ref{app:subsec3} and~\ref{app:subsec4}, respectively.
\subsection{The amplitudes\label{sec:3.3}}
Let $\bm{l}=(N_L,N_R,C_L,C_R,C_L^1,C_R^1,C_L^2,C_R^2,g,r,t_2t_1t_0)$ be an 11-tuple property for a direction sequence beginning with $RL$, where $N_X$ ($X\in\{L,R\}$) is the number of $X$'s, $C_X$ is the number of $X$ clusters, $C_X^1$ is the number of $X$ clusters of size one, $C_X^2$ is the number of $X$ clusters of size two, $g$ is the number of $S$ and $M$ groups, $r$ is the number of singular $R$ clusters bordered by a singular $L$ cluster and a non-singular $L$ cluster, and $t_2t_1t_0$ indicates the type of the end of the sequence. By expression~(\ref{eq:13}), the phases of all direction sequences with the same $\bm{l}$ are identical. Thus, by counting the number $cnt_{\bm{l}}$ of direction sequences that possess the property $\bm{l}$ and that lead to a specific final state, and then summing over $cnt_{\bm{l}}$ for all possible values of $\bm{l}$, one can obtain the amplitude for that state. To simplify the amplitude expression, we introduce three new combinatorial symbols as follows.
\begin{myDefinition}\label{def:def3}
The number of compositions of the integer $u$ into $m$ parts in which there are $v$ summands 2 but no summand 1 is
\begin{equation}
\begin{pmatrix}
u \\ m \\ v
\end{pmatrix}
\equiv
\begin{cases}
\dbinom{m}{v}\dbinom{u-2m-1}{m-v-1},&\begin{matrix}\text{if }m>v\text{ and } u\ge 3m-v\end{matrix}\\
1,&\text{otherwise}
\end{cases}
\end{equation}
\end{myDefinition}
One can verify that there is exactly one composition~\cite{Merlini2004Compositions} for $m=v=u/2$; however, if $m<v$ or $u<3m-v$, then the required composition would not exist at all. We define $\bigl(\begin{smallmatrix}u\\m\\v\end{smallmatrix}\bigr)$ to be 1 rather than 0 for this impossible case so that it can be generally applied in the subsequent amplitude formula without introducing any invalid expression.
\begin{myDefinition}\label{def:def4}
Consider a union of two sets $U=A\bigcup B$, where $A$ contains $x$ elements and $B$ contains $u-x$ elements. The number of $m$-subsets $C$ of $U$ in which $r$ of the $m$ elements of $C$ are taken from $A$, and the remaining $m-r$ elements of $C$ are taken from $B$ is
\begin{equation}
\begin{pmatrix}
u \\ m \\ x \\ r
\end{pmatrix}
\equiv
\begin{cases}
\dbinom{x}{r}\dbinom{u-x}{m-r},&\begin{matrix}\text{ if }0\leq r\leq x\text{ and } 0\leq m-r\leq u-x\end{matrix}\\
0,&\text{\ \,otherwise}
\end{cases}
\end{equation}
\end{myDefinition}
\begin{myDefinition}\label{def:def5}
The number of ways to construct a permutation of the multi-set $\{x\cdot S, y\cdot M\}$ where the number of $S$ and $M$ groups is $g$, and in which the last element is indicated by $t_1t_0$ is
\begin{equation}
\widetilde{\begin{pmatrix}
x \\ y \\ g \\ t_1t_0
\end{pmatrix}}
\equiv
\begin{cases}
t_0\tbinom{x-1}{\lfloor \tfrac{g}{2} \rfloor-1}\tbinom{y-1}{\lceil \tfrac{g}{2} \rceil-1}+t_1\tbinom{x-1}{\lceil \tfrac{g}{2} \rceil-1}\tbinom{y-1}{\lfloor \tfrac{g}{2} \rfloor -1},\text{ if }g\ge 2\text{ and}\\
\qquad g \leq 2\text{min}(x,y)+t_1\mathbf{1}_{\mathbb{Z}^+}(x-y)+t_0\mathbf{1}_{\mathbb{Z}^+}(y-x)\\
\tbinom{x-1}{\lfloor \tfrac{g}{2} \rfloor-1}\tbinom{y-1}{\lceil \tfrac{g}{2} \rceil-1},\text{ if }\lfloor\tfrac{g}{2}\rfloor\leq x<\lceil\tfrac{g}{2}\rceil\leq y\\
\tbinom{x-1}{\lceil \tfrac{g}{2} \rceil-1}\tbinom{y-1}{\lfloor \tfrac{g}{2} \rfloor-1},\text{ if }\lfloor\tfrac{g}{2}\rfloor\leq y<\lceil\tfrac{g}{2}\rceil\leq x\\
1,\text{ if } xy=0, x+y>0, \text{ and } g=1\\
0,\text{ otherwise}
\end{cases}
\end{equation}
\end{myDefinition}
Note that such a multi-set permutation~\cite{Brualdi2009combinatorics} corresponds to the $L$ cluster mask for a direction sequence with $I$'s eliminated. The derivation of the expressions on the right is presented in Appendix~\ref{app:subsec5}. With these new symbols, a general amplitude expression is demonstrated in Theorem~\ref{theo:theo1}, which is proved in appendix~\ref{app:subsec6}.
\begin{myTheorem}[A general formula for the amplitudes\label{theo:theo1}]In the $n$-step ($n=N_R+N_L-2$) order-3 quantum walk beginning with the state $\Ket{0,1,0,0}$ and terminating at the position $k$ ($k=N_R-N_L$), the amplitude of a final state can be generally formulated as
        \begin{equation}
        \sum_{\substack{C_L,C_R,C_L^1,C_R^1,\\C_L^2,C_R^2,g,r}}\frac{(-1)^{n+C_L+C_R+C_L^2+C_R^2+r}}{\sqrt{2^n}}\rho\cdot cnt_{\bm{l}}
        \end{equation}
where
\begin{equation*}
cnt_{\bm{l}}=
\widetilde{\begin{pmatrix}
        C_L^1\!-\!t_2 \\ C_L\!-\!C_L^1 \\ g \\ t_1t_0
        \end{pmatrix}}\!
        \begin{pmatrix}
        C_R\!-\!1 \\ C_R^1\!-\!1 \\ g\!+\!t_1\!\cdot\!t_2\!-\!1 \\ r
        \end{pmatrix}\!
        \begin{pmatrix}
        N_L\!-\!C_L^1 \\ C_L\!-\!C_L^1 \\ C_L^2
        \end{pmatrix}\!
        \begin{pmatrix}
        N_R\!-\!C_R^1 \\ C_R\!-\!C_R^1 \\ C_R^2
        \end{pmatrix},
\end{equation*}
 and $\rho$ is the fraction of direction sequences leading to the final state.
\end{myTheorem}

By restricting the general bounds of $C_L,C_R,C_L^1,C_R^1,$ $C_L^2,C_R^2,g$ and $r$ to the ranges of the corresponding properties of sequences leading to a specific final state and separating the extreme cases out from the summations, the previous expression can be stated more concretely. Due to length constraint, we put the detailed derivation for each amplitude $a_j$ ($j\in\{0,1,\dots,7\}$) in the full version of this paper~\cite{Zhou2019QW2M}.
\section{Numerical simulations\label{sec:4}}
\begin{figure}
\includegraphics{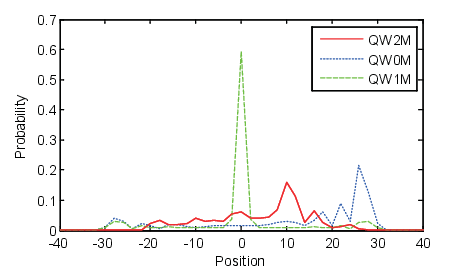}
\caption{\label{fig:fig1} (color online) Comparisons of three probability distributions with single-component initial states. }
\end{figure}

\begin{figure}
\includegraphics{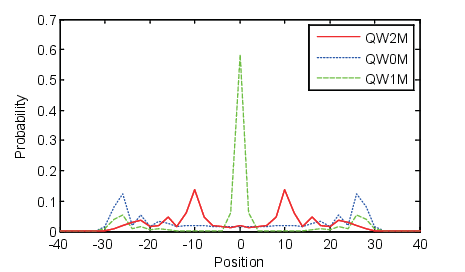}
\caption{\label{fig:fig2} (color online) Comparisons of three probability distributions with symmetric initial states.}
\end{figure}

We show in Fig~\ref{fig:fig1} and Fig.~\ref{fig:fig2} the probability distributions after 40 steps for the three kinds of walks (QW0M, QW1M, and QW2M) with Hadamard coin, where only even positions are plotted since the probabilities are all zero for odd positions. The simulation generating these distributions is carried out with MATLAB, and the commented simulation code of QW2M is included in the full version of this paper~\cite{Zhou2019QW2M}.

In Fig.~\ref{fig:fig1}, the initial states for QW0M, QW1M and QW2M are $\Ket{0,1}$, $\Ket{1,0,0}$, and $\Ket{0,1,0,0}$, respectively (so that the three distributions have the same bias), while in Fig.~\ref{fig:fig2}, the initial states for the three cases are $(\Ket{0,0}+i\Ket{0,1})/\sqrt{2}$, $(\Ket{1,0,0}+i\Ket{1,0,1}+\Ket{-1,0,0}+i\Ket{-1,0,1})/2$, and $(1/2\sqrt{2})(\Ket{0,1,0,0}+i\Ket{0,1,0,1}+\Ket{0,-1,0,0}+i\Ket{0,-1,0,1}+\Ket{2,1,0,0}+i\Ket{2,1,0,1}+\Ket{-2,-1,0,0}+i\Ket{-2,-1,0,1})$, respectively. These figures indicate that the localization property of QW1M, a single peak at the origin, does not appear in either of the two distributions of QW2M;  instead, a biased peak and two symmetric peaks near positions $\pm 10$ occur, respectively. Additionally, the symmetric distribution of QW2M shows oscillatory behaviour beyond the two peaks, while displays a relatively smooth curve between the two peaks.
\section{Conclusion\label{sec:5}}
In this paper, we concretely defined a Hadamard quantum walk with two-step memory, numerically studied its distributions and analytically derived a general formula for its amplitudes. Although this walk is an extension of the Hadamard QW1M, it does not possess the localization property that QW1M has, while exhibits some similarities with the normal Hadamard quantum walk without memory (e.g., two symmetric peaks on both sides of the origin, oscillatory behaviour, and smooth distribution in the middle). After simulating QW2M for various number $n$ of steps, we observed that the positions of the two peaks are roughly $\pm n/4$, which are much nearer to the origin than that ($\pm n/\sqrt{2}$) in QW0M, and the probabilities beyond the two peaks in QW2M are not negligible. In addition, the general form we developed for the amplitudes of QW2M indicates that as the number of historical steps in memory gets large, the amplitude expression for each final state gets increasingly complicated.

Other models of quantum walks with history dependence have been proposed by using multiple coins or mixing coin~\cite{Flitney2003History,Gettrick2014Cycles,Gettrick2016Graphs,Rohde2013recycled,Stang2009Correlation}. The distribution for 3-coin history-dependent quantum walk (HDQW for short)~\cite{Flitney2003History} seems close in shape to our result. By performing a 100-step simulation (same steps as that in~\cite{Flitney2003History}), we observe that the two peaks in QW2M (near the positions $\pm 25$) are farther to the origin than that in HDQW (near the positions $\pm 20$). Moreover, the probabilities at the two peaks exceed 0.1 in QW2M, while they are less than 0.1 in HDQW. The long-term-memory walk introduced by Rohde et al.~\cite{Rohde2013recycled} keeps history information with recycled coins, whose state space is isomorphic to that of QW2M. It performs memory-dependent coin flip operator on different coin states (called memory elements) during a cycle and gives a binomial distribution (including two-step memory, which corresponds to $N=3$ memory elements); while QW2M always performs the coin-flip transform on the same coin state, and yields a double-peaked distribution.
%


%
\appendix
\section{Appendix : Proofs of the lemmas and the theorem\label{app:appA}}
\subsection{\label{app:subsec1}Proof of Lemma 1}
A long $L$ or $R$ cluster of size $j$ $(j\ge 4)$ gives a phase contribution of $(-1)^{j-3}$, and a cluster of size fewer than four gives no contribution (on its own). If there are two long $L$ clusters (of size greater than three), say, $C$ and $C'$ , one can move one $L$ from $C$ to $C'$ until the length of $C$ shrinks to three, without changing the combined phase contribution. To obtain the total contribution of all long $L$ clusters in a direction sequence, one can select a long $L$ cluster as a target $C_\text{target}$, then repeatedly move $L$'s from all long $L$ clusters but the target one to $C_\text{target}$ until no clusters except $C_\text{target}$ contain more than three elements. After these moves, the size of $C_\text{target}$ implies the total phase contribution of all long $L$ clusters in the original direction sequence.

Suppose there are $N_L$ $L$'s and $C_L$ $L$ clusters in the original sequence, in which the number of $L$ clusters of size one, of size two, of size three and of size greater than three are $C_L^1$, $C_L^2$, $C_L^3$, and $C_L^{4+}$, respectively, then we have $C_L^1+C_L^2+C_L^3+C_L^{4+}=C_L$. After the moves, the size of $C_\text{target}$ is $N_L-\lceil C_L^1+2C_L^2+3(C_L^3+C_L^{4+}-1)\rceil=N_L+2C_L^1+C_L^2-3C_L+3$, which gives a contribution of $(-1)^{N_L+2C_L^1+C_L^2-3C_L+3-3}=(-1)^{N_L+C_L^2+C_L}$. Similarly, the contribution given by all long $R$ clusters is $(-1)^{N_R+C_R^2+C_R}$, so the total contribution of long clusters is $(-1)^{n+C_L+C_R+C_L^2+C_R^2}$.
\subsection{\label{app:subsec2}Proof of Lemma 2}
Apart from long clusters, a phase contribution of $-1$ occurs once a singular cluster comes after another singular cluster, and it increases to $(-1)^2$ if a third singular cluster comes, while stays unchanged if the third cluster is a non-singular one. In other words, a string of singular clusters $\overline{S}S$ (or $S\overline{S}$) gives a contribution of -1, and $\overline{S}S\overline{S}$ (or $S\overline{S}S$) gives a contribution of $(-1)^2$. An induction on the size of the string of singular clusters indicates that a string of singular clusters $\dots \overline{S}S\overline{S}S\dots$ of size $m$ (bordered by non-singular clusters) gives a phase contribution of $(-1)^{m-1}$.

Also, one can state the contribution of singular clusters in another way. First, we define the phase contribution of an $\overline{S}$ to be $(-1)^x$, where $x\in\{0,1,2\}$ is the number of $S$'s bordering this $\overline{S}$. For example, in $M\overline{S}S\overline{S}S\overline{M}M\overline{S}I$, the first $\overline{S}$ gives a contribution of $(-1)^1$, the second $\overline{S}$ gives a contribution of $(-1)^2$, and the last $\overline{S}$ gives a contribution of $(-1)^0$. Then, the phase contribution of a string of singular clusters is equivalent to the product of the contributions given by all $\overline{S}$'s within the string, which holds for multiple strings of singular clusters as well as a complete direction sequence.

Let $r$ be the number of $\overline{S}$'s bordered by an $S$ and an $M$ or by an $S$ and an $I$ (the order does not matter) in the cluster mask for a direction sequence, and let $r'$ be the number of $\overline{S}$'s bordered by two $S$'s. Then, the phase contribution given by all $\overline{S}$'s in the mask is $(-1)^{r+2r'+0}=(-1)^r$, which is also the contribution of all alternate direction segments in the direction sequence.
\subsection{\label{app:subsec3}Proof of Lemma 3}
Here we only deduce the range of $C_L^2$, the range of $C_R^2$ can be obtained in similar. If $C_L^1=C_L=N_L$, then $C_L^2=0=C_L-C_L^1=N_L-C_L$. For $0\leq C_L^1\leq C_L-1$, let $C_L^M=C_L-C_L^1\ge1$ be the number of multi-element $L$ clusters in the direction sequence, then $N_L-C_L^1\ge 2C_L^M$, which can be divided into three cases.

If $N_L-C_L^1=2C_L^M$, then each multi-element $L$ cluster contains two $L$'s exactly, namely, $C_L^M=C_L^2=C_L-C_L^1=(N_L-C_L^1)/2$, which gives $C_L^1=2C_L-N_L$ and $C_L^2=C_L-C_L^1=N_L-C_L$.

If $2C_L^M<N_L-C_L^1<3C_L^M$, then the lower bound of $C_L^2$ is $b_\text{min}=3C_L^M-(N_L-C_L^1)=3C_L-2C_L^1-N_L$, which can be achieved by repeatedly moving one $L$ from an $L$ cluster of size greater than three to an $L$ cluster of size two, until there does not exist any $L$ cluster of size greater than three. From the sequence in which $C_L^2=b_\text{min}$, one can progressively increase the number of $L$ clusters of size two as follows: first, choose an $L$ cluster of size three to be a moving target $C_\text{target}$, then repeatedly move one $L$ from an $L$ cluster of size three (but not $C_\text{target}$) to $C_\text{target}$, until $C_\text{target}$ is the only one $L$ cluster of size greater than two. During such a process, each move induces an increment of $C_L^2$, and $C_L^2$ reaches the upperbound $b_\text{max}=C_L^M-1=C_L-C_L^1-1$ eventually.

If $N_L-C_L^1\ge 3C_L^M$, then $b_\text{min}$ is negative, and the number of $L$ clusters of size two is zero (the new lowerbound) when each multi-element $L$ cluster contains more than two $L$'s. In addition, from the lowerbound 0, one can progressively increase $C_L^2$ until it reaches $b_\text{max}$.

Therefore, either $C_L^2=C_L-C_L^1=N_L-C_L$ (along with $C_L^1=2C_L-N_L$), or $C_L^2$ varies from $\text{max}(0,3C_L-2C_L^1-N_L)$ to $C_L-C_L^1-1$ (along with $\text{max}(0,2C_L-N_L+1)\leq C_L^1\leq C_L-1$).
\subsection{\label{app:subsec4}Proof of Lemma 4}
The required cluster mask can be constructed as follows: 1) give a permutation of the multiset $\{(C_L^1-t_2)\cdot S,(C_L-C_L^1)\cdot M\}$, then append an $I$ at the end of the permutation if $t_2=1$, so that $m_L$ is set up; 2) put $C_R^1$ $\overline{S}$'s in $C_R^1$ of the $C_R$ places (for $R$ clusters) between and after the elements of $m_L$, and if the last $\overline{S}$ is after the last $L$ cluster (an $S$ or an $M$), replace it with $\overline{I}$; 3) put $C_R-C_R^1$ $\overline{M}$'s in the remaining places for $R$ clusters. In the second step, once an $\overline{S}$ is placed between an $S$ and an $M$, or between an $S$ and an $I$, $r$ is incremented by one. For convenience, we refer to the places that induce an increment of $r$ when an $R$ cluster is put in as \emph{positive places}, and refer to the remaining places for $R$ clusters as \emph{negative places}.

We first consider the range of $g$. In $m_L$, the number of $S$ groups and $M$ groups reaches its minimum $g_\text{min}$ when all $S$'s form a single $S$ group and all $M$'s form a single $M$ group, and $g$ reaches its maximum $g_\text{max}$ if each $S$ or $M$ group contains a single element. Let $C_L^S=C_L^1-t_2$ and $C_L^M=C_L-C_L^1$ respectively denote the number of $S$'s and the number of $M$'s in $m_L$. Then, $g_\text{min}$ and $g_\text{max}$ should be deduced for four different cases: (1) When $C_L^SC_L^M\ne 0$ and $C_L^S\ne C_L^M$, we have $g_\text{min}=2$, and $g_\text{max}$ could take two distinct values: a) If $C_L^M>C_L^S$ and $t_0=1$, or $C_L^S>C_L^M$ and $t_1=1$, or $t_2t_1t_0=011$, then $g_\text{max}=2\text{min}(C_L^S,C_L^M)+1$; b) otherwise, $g_\text{max}=2\text{min}(C_L^S,C_L^M)$. (2) When $C_L^S=C_L^M\ne 0$, we have $g_\text{min}=2$ and $g_\text{max}=2C_L^S$ for any $t_2t_1t_0$. (3) When $C_L^S>C_L^M=0$ ($t_1t_0$ must be $10$), or $C_L^M>C_L^S=0$ ($t_1t_0$ must be $01$), we have $g_\text{min}=g_\text{max}=1$. (4) When $C_L^S=C_L^M=0$, we have $g_\text{min}=g_\text{max}=0$. Thus, the range of $g$ can be formulated as
$2-\delta_{C_L^S,0}-\delta_{C_L^M,0}\leq g\leq 2\text{min}(C_L^S,C_L^M)+t_1\mathbf{1}_{\mathbb{Z}^+}(C_L^S-C_L^M)+t_0\mathbf{1}_{\mathbb{Z}^+}(C_L^M-C_L^S)$
for all cases, and the expression (\ref{eq:16}) follows.

The bounds of $r$ rely on the number of positive places, which can be expressed in terms of $g$ and $t_1\cdot t_2$: if $t_1\cdot t_2=1$ (i.e., $t_1=t_2=1$), then there are $g$ positive places; and if $t_1\cdot t_2=0$, then there are $g-1$ positive places; thus, the number of positive places in both cases is $g+t_1\cdot t_2-1$. Since there are $C_R$ places (for $R$ clusters) in total, the number of negative places is $C_R-g-t_1\cdot t_2+1$. To maximize $r$, the $R$ clusters of size one need to be put in positive places as many as possible, so the upper bound of $r$ is $r_\text{max}=\text{min}(C_R^1-1,g+t_1\cdot t_2-1)$ (there is one $R$ cluster of size one---the first $R$ in the sequence---has occupied a negative place). To minimize $r$, the $R$ clusters of size one should be put in negative places as many as possible, thus the lower bound of $r$ is $r_\text{min}=\text{max}(0,C_R^1-C_R+g+t_1\cdot t_2-1)$. Note that from $r_\text{max}$, one can progressively decrease $r$ by moving one $R$ cluster of size one from a positive place to a negative place, meaning that $r$ can take any integer value between (inclusive) $r_\text{min}$ and $r_\text{max}$.
\subsection{\label{app:subsec5}Derivation for the expression in Definition 5}
Let $g_S$ and $g_M$ respectively be the number of $S$ groups and the number of $M$ groups in the required permutation $P_M$, then $g_S+g_M=g$, and $P_M$ corresponds to a composition of the integer $x$ into $g_S$ parts followed by a composition of the integer $y$ into $g_M$ parts. Since $S$ groups and $M$ groups appear alternately, we have $g_S=\lfloor g/2\rfloor$ and $g_M=\lceil g/2\rceil$, or $g_S=\lceil g/2\rceil$ and $g_M=\lfloor g/2\rfloor$ if $g$ is odd, and $g_S\!=\!g_M\!=\!g/2\!=\!\lfloor g/2\rfloor\!=\!\lceil g/2\rceil$ if $g$ is even.

We first consider the case that $2\leq g\leq 2\text{min}(x,y)+t_1\mathbf{1}_{\mathbb{Z}^+}(x-y)+t_0\mathbf{1}_{\mathbb{Z}^+}(y-x)$. If the last element of $P_M$ is $S$ ($t_1t_0=10$), then $g_S\ge g_M$, thus $g_S=\lceil g/2\rceil$ and $g_M=\lfloor g/2\rfloor$, and the number of ways to construct $P_M$ is $\tbinom{x-1}{g_S-1}\tbinom{y-1}{g_M-1}=t_1\tbinom{x-1}{\lceil g/2\rceil-1}\tbinom{y-1}{\lfloor g/2\rfloor-1}$.
In similar, if the last element is $M$ ($t_1t_0=01$), then we have $g_S\leq g_M$, thus $g_S=\lfloor g/2\rfloor$ and $g_M=\lceil g/2\rceil$, and the number of ways to construct $P_M$ is $t_0\tbinom{x-1}{\lfloor g/2\rfloor -1}\tbinom{y-1}{\lceil g/2\rceil-1}$.

When the last element of $P_M$ is not specified ($t_1t_0=11$), we need to count in two cases---$P_M$ ends with $M$ or $S$. If $g$ is odd, then $g_S=\lfloor g/2\rfloor$ and $g_M=\lceil g/2\rceil$, or $g_S=\lceil g/2\rceil$ and $g_M=\lfloor g/2\rfloor$, and the positions of $S$ groups and $M$ groups are deterministic for either case, so the number of ways to construct $P_M$ is
\begin{equation}
t_0\dbinom{x\!-\!1}{\lfloor \tfrac{g}{2}\rfloor\!-\!1}\dbinom{y\!-\!1}{\lceil \tfrac{g}{2}\rceil\!-\!1}+t_1\dbinom{x\!-\!1}{\lceil \tfrac{g}{2}\rceil\!-\!1}\dbinom{y\!-\!1}{\lfloor \tfrac{g}{2}\rfloor\!-\!1}.\label{eq:21}
\end{equation}
While if $g$ is even, then $g_S=g_M=g/2$, and the first $S$ group could be placed before or after the first $G$ group, so the number of ways to construct $P_M$ is $2\tbinom{x-1}{g/2-1}\tbinom{y-1}{g/2-1}$,
which equals~(\ref{eq:21}).

As a result, expression~(\ref{eq:21}) applies for any $g$ (integer) satisfying $2\leq g\leq 2\text{min}(x,y)+t_1\mathbf{1}_{\mathbb{Z}^+}(x-y)+t_0\mathbf{1}_{\mathbb{Z}^+}(y-x)$.

It remains to consider the four extreme cases. 1) If $g$ is odd and $\lfloor g/2\rfloor \leq x<\lceil g/2\rceil \leq y$, then $g_S=\lfloor g/2 \rfloor$ and $g_M=\lceil g/2 \rceil$ (for sure), and the first group is an $M$ group. 2) Similarly, if $\lfloor g/2\rfloor \leq y<\lceil g/2\rceil \leq x$, then $g_M=\lfloor g/2 \rfloor$ and $g_S=\lceil g/2 \rceil$, and the first group is an $S$ group. 3) If $x>y=0$ or $y>x=0$, then there is only one group (i.e., $g=1$), which result in a single permutation. 4) If $g$ is out of the valid range, then the required permutation becomes impossible, so the number of permutation is 0.
\subsection{\label{app:subsec6}Proof of Theorem 1}
Let $cnt_{\bm{l}}$ be the number of direction sequences beginning with $RL$ and characterized by $\bm{l}=(N_L,N_R,C_L,C_R$, $C_L^1,C_R^1,C_L^2,C_R^2,g,r,t_2t_1t_0)$, then $cnt_{\bm{l}}$ can be viewed as the number of ways to construct a direction sequence with property $\bm{l}$, which can be achieved by the following four steps.
\begin{enumerate}[(1)]
\item construct a permutation of the multi-set $\{(C_L^1-t_2)\cdot S, (C_L-C_L^1)\cdot M\}$ such that the number of $S$ and $M$ groups is $g$, and that the last element is indicated by $t_1t_0$; then put an $I$ at the end of the permutation if $t_2=1$. With the combinatorial symbol defined in Definition~\ref{def:def5}, the number of ways to do this is
    \begin{equation}
    \setlength\abovedisplayskip{3pt}
    \setlength\belowdisplayskip{1pt}
    N^{(1)}=
    \widetilde{
    \begin{pmatrix}
    C_L^1-t_2 \\ C_L-C_L^1 \\ g \\ t_1t_0
    \end{pmatrix}}.
    \end{equation}
\item Replace each $S$ and each $I$ with an $L$ cluster of size one, then replace each $M$ with a multi-element $L$ cluster; among those multi-element clusters there are $C_L^2$ clusters of size two. This corresponds to a composition of the integer $N_L-C_L^1$ into $C_L-C_L^1$ parts such that there are $C_L^2$ summands 2 but no summand 1, which can be done in
    \begin{equation}
    \setlength\abovedisplayskip{3pt}
    \setlength\belowdisplayskip{1pt}
    N^{(2)}=
    \begin{pmatrix}
    N_L-C_L^1 \\ C_L-C_L^1 \\ C_L^2
    \end{pmatrix}
    \end{equation}
    ways.
\item Put $r$ $R$'s in $r$ of the $g-t_1\cdot t_2-1$ positive places, add an $R$ before the first $L$ (which has occupied a negative place), and then put the remaining $R$ clusters of size one in $C_R^1-r-1$ of the $C_R-g-t_1\cdot t_2$ negative places. This can be accomplished in $\tbinom{g+t_1\cdot t_2-1}{r}\tbinom{C_R-g-t_1\cdot t_2}{C_R^1-r-1}$ ways, namely,
    \begin{equation}
    \setlength\abovedisplayskip{1pt}
    \setlength\belowdisplayskip{1pt}
    N^{(3)}=
    \begin{pmatrix}
    C_R-1 \\ C_R^1-1 \\ g+t_1\cdot t_2-1 \\ r
    \end{pmatrix}.
    \end{equation}
\item  Put $C_R-C_R^1$ multi-element $R$ clusters (involving $N_R-C_R^1$ $R$'s) in the remaining places for $R$ clusters, $C_R^2$ of which are of size two. The number of ways to do this is
    \begin{equation}
    \setlength\abovedisplayskip{1pt}
    \setlength\belowdisplayskip{1pt}
    N^{(4)}=
    \begin{pmatrix}
    N_R-C_R^1 \\ C_R-C_R^1 \\ C_R^2
    \end{pmatrix}.
    \end{equation}
\end{enumerate}
Hence the number of direction sequences characterized by $\bm{l}$ is $cnt_{\bm{l}}=\begin{matrix}\prod_{i=1}^4 N^{(i)}\end{matrix}$.

Suppose $\rho \cdot cnt_{\bm{l}}$ of these sequences leads to the final quantum state $\Ket{n_3,n_2,k,p}$, then summing $\rho \cdot cnt_{\bm{l}}$ over all possible values of $C_L,C_R,C_L^1,C_R^1,C_L^2,C_R^2,g$ and $r$ (restricted according to $\Ket{n_3,n_2,k,p}$) gives the amplitude of $\Ket{n_3,n_2,k,p}$. Therefore, Theorem~\ref{theo:theo1} gives a general form of the quantum amplitudes of Hadamard QW2M.
\end{document}